\documentclass[iop]{emulateapj}

\begin{document}


\title{Excess C/O and C/H in outer protoplanetary disk gas}



\author{Karin I. \"Oberg}
\affil{Harvard-Smithsonian Center for Astrophysics, 60 Garden St., Cambridge, MA 02138, USA}

\and

\author{Edwin A. Bergin}
\affil{Department of Astronomy, University of Michigan, 311 West Hall, 1085 S. University Ave, Ann Arbor, MI 48109, USA}

\begin{abstract}

The compositions of nascent planets depend on the compositions of their birth disks. In particular, the elemental compositions of Gas Giant gaseous envelopes depend on the elemental composition of the disk gas from which the envelope is accreted. Previous models demonstrated that sequential freeze-out of O and C-bearing volatiles in disks will result in an supersolar C/O ratios and subsolar C/H ratios in the gas between water and CO snowlines. This result does not take into account, however, the expected grain growth and radial drift of pebbles in disks, and the accompanying re-distribution of volatiles from the outer to the inner disk. Using a toy model we demonstrate that when drift is considered, CO is enhanced between the water and CO snowline, resulting in both supersolar C/O and C/H ratios in the disk gas in the Gas Giant formation zone. This result appears robust to the details of the disk model as long as there is substantial pebble drift across the CO snowline, and the efficiency of CO vapor diffusion is limited. Gas Giants that accrete their gaseous envelopes exterior to the water snowline and do not experience substantial core-envelope mixing, may thus present both superstellar C/O and C/H ratios in their atmospheres. Pebble drift will also affect the nitrogen and noble gas abundances in the planet forming zones, which may explain some of Jupiter's peculiar abundance patterns.

\end{abstract}

\keywords{astrochemistry --- protoplanetary disks  ---  molecular processes --- planet-disk interactions --- planets and satellites: atmospheres --- planets and satellites: formation}

\section{Introduction  \label{sec:intro}}

Gas Giants form in protoplanetary disks, through core accretion followed by runaway gas accretion, or through gravitational instabilities \citep{Lissauer87,Boss97,Boley09}.  In the core accretion scenario, the elemental compositions of Gas Giant gaseous envelopes or atmospheres are determined by the composition of the disk gas from which the envelope is accreted, by subsequent accretion of icy planetesimals, and by core-envelope interactions \citep{Lodders02,Hersant04,Madhusudhan14}. In this letter we show how snowlines in conjunction with pebble drift affect the C/O and C/H ratios of disk gas, and further the compositions of of Gas Giant atmospheres forming through core accretion.

Considering a core accretion idealized scenario, where the core is composed purely of disk solids and the envelope purely from disk gas, the gaseous envelope will have a high C/O ratio and a substellar C/H ratio if it is accreted in between the water and CO snowlines \citep{Oberg11e}. At the water and CO$_2$ snowlines the gas is (preferentially) depleted of oxygen, resulting in a supersolar gas-phase C/O ratio and subsolar O/H and C/H ratios. This motivated \citet{Oberg11e} to predict that Gas Giants that form through the core accretion scenario at large separations should have a C/O ratio close to unity in their envelopes, and a substellar C/H ratio. In reality pollution by pebbles and planetesimals accreting with the gas or at later times is bound to take place, and this process and migration will affect the elemental atmospheric ratios (see \S4). 

Both the C/O ratio and the C/H ratio (or metallicity) have large effects on the atmospheric chemistry of Gas Giants \citep{Fortney08,Lodders10,Madhusudhan12,Moses13}. Observations of Gas Giant atmospheric chemical compositions may thus be used to constrain their elemental composition, and further their formation mode and location. In the Solar System, the atmospheres of both Jupiter and Saturn appear to be enhanced in C/H (and N/H) relative to Solar by factors of a few \citep{Owen99,Wong04,Atreya05}. The C/O ratio remains in question in the Solar System Gas Giants, but may become resolved when Juno measures the water content of Jupiter \citep{Helled14}. The C/O ratio has been constrained in  a number of Gas Giant exoplanet atmospheres and several appear to contain supersolar C/O \citep{Madhusudhan11,Moses13,Lee13}, though some of the early results are contested \citep[e.g.][]{Kreidberg15}. The metallicity of a few extrasolar Gas Giants have been retrieved as well, and the result is a range of inferred C/H ratios \citep{Madhusudhan11,Moses13, Lee13}.  Of especial interest for this letter is the retrieval of a C/O ratio close to unity and a supersolar C/H in the large-separation Gas Giants HR 8799b and c\citep{Lee13,Lavie16}, which is at odds with the predictions from \citet{Oberg11e}.

\begin{figure*}[ht!]
\figurenum{1}
\begin{centering}
\includegraphics[width=0.7\textwidth]{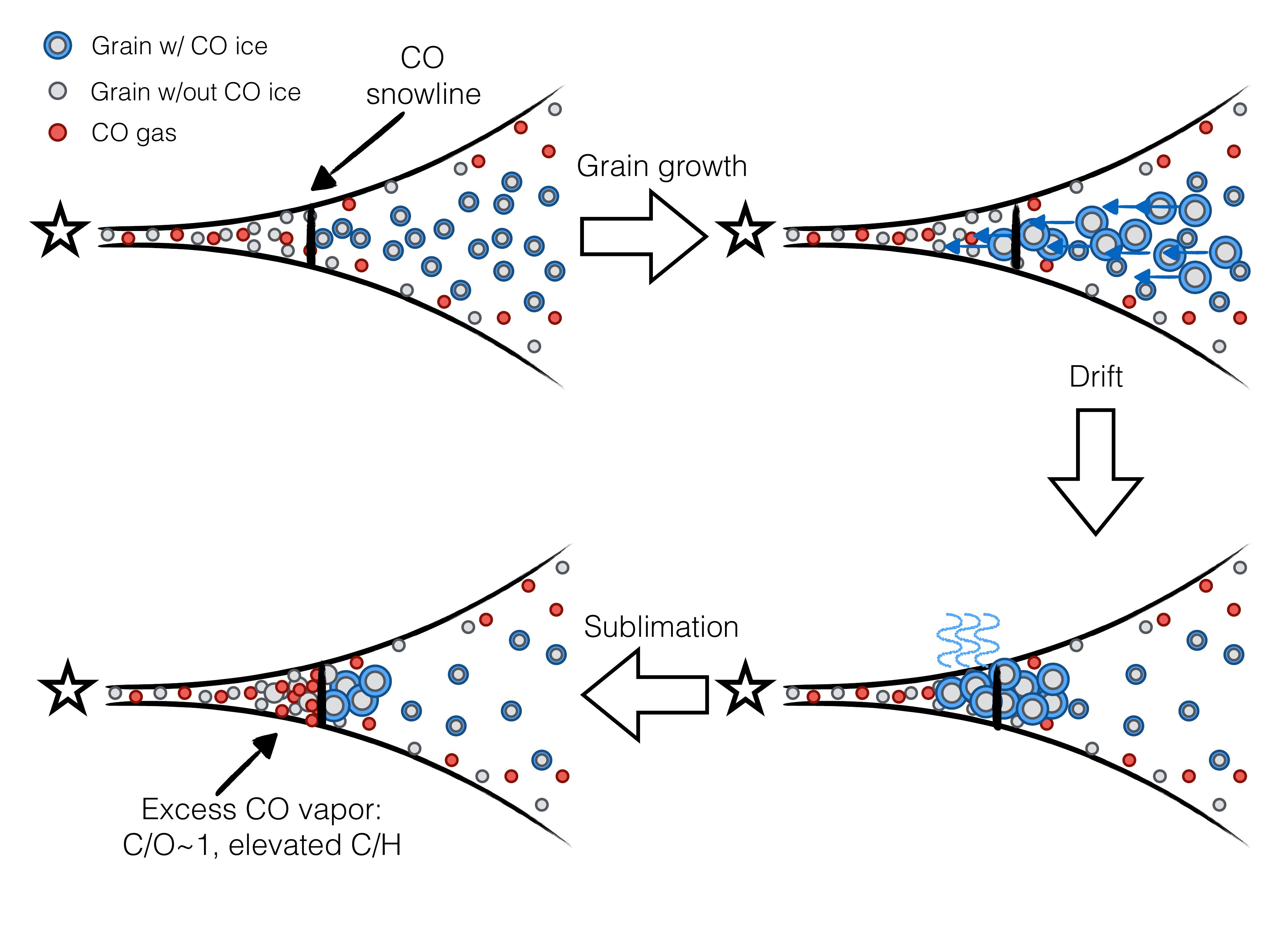}
\caption{Cartoon view of the model framework focusing on the region around the CO snowline. Initially the gas and grains are co-located and the CO gas abundance with respect to H$_2$ is determined by a balance between freeze-out and sublimation. As grains coagulate and grow to form pebbles, they begin to drift. CO-ice-covered pebbles that drift across the CO snowline will sublimate, locally increasing the gas-phase CO abundance and producing a region where the gas-phase C/O ratio is $\sim$1 and C/H is enhanced compared to a static disk. \label{fig:cartoon}}
\end{centering}
\end{figure*}

The prediction of substellar gas-phase C/H in the outer disk was based on a static disk model, however, while in reality disks are dynamical systems that experiences large-scale redistributions of gas and solids. In this letter we explore how the redistribution of volatiles that accompanies grain growth and subsequent pebble drift affect the elemental composition of the gas in the planet forming zone of disks. We use a simple toy model to demonstrate the importance of this effect and present a new set of qualitative predictions on the C/O and C/H ratios in Gas Giants assembling at different disk radii.

\section{Model framework\label{sec:model}} 

The framework for our toy model is a disk where the distribution of solids is regulated by settling, growth and drift. Grain growth through coagulation is efficient up to cm-scales or pebble sizes and takes place on time scales of thousands of years at 1~AU \citep{Weidenschilling80,Nakagawa81,Dullemond05,Birnstiel10}. In the outer disk, grain growth occurs on longer time scales, tens of thousands of years, but this is still short compared to disk lifetimes \citep{Brauer08}. 
Once formed, pebbles are expected to rapidly drift inward due to gas drag, resulting in a major redistribution of solids from the outer to the inner disk. This scenario is confirmed by observations of disks with mm-to-cm sized pebbles confined to the inner 10s of AU, while gas and small, micro-sized dust grains extend to 100s of AU \citep{Wilner05,Andrews12,Perez12,Rosenfeld13,Perez15,Tazzari16}. Why drift of pebbles stops or slows down at 10s of AUs in these disks is not immediately clear, but may be due to the presence of sub-structure \citep[e.g.][]{Brogan15,Andrews16}. 

Based on these theoretical and observational constraints, and on the insights from \citet{Oberg11e}, we imagine the scenario outlined in cartoon-form in Fig. \ref{fig:cartoon}. The disk forms with gas and solids (in the form of bare and icy dust grains) initially co-located. The dust grain composition is set by sublimation and condensation rates of different volatiles, which are temperature and therefore disk radius dependent. Beyond the water snowline, the grains are covered by water ice, and further outward also by CO$_2$ and CO ice. In between the water and CO snowlines the gas is depleted in oxygen with respect to carbon, resulting in a supersolar C/O ratio. As the icy grains coagulate to form pebbles they drift rapidly inward \citep{Birnstiel10}, depleting the outer disk of solids. Pebbles that form exterior to the CO snowline, which is located at 20-60~AU around Solar-type stars \citep{Qi13c,Piso15,Schwarz16}, and drift inward deplete the outer disk of CO ice \citep{Bergin16} as well as more refractory solids. If CO-ice covered pebbles drift across the CO snowline and sublimate some distance interior to it \citep{Piso15}, the inner disk will become enhanced in CO vapor, locally enhancing the C/H (and O/H) ratio above the expected value for a static disk. In this letter we only consider the effects of CO redistribution in detail, but the process can be generalized to other volatiles. In particular, icy pebbles that drift past the CO$_2$ snowline will result in a second disk region with a C/H and C/O excess in the gas. 

Following sublimation, CO will be distributed over some disk region due to diffusion.  Diffusion has been shown to have a large effect on elemental ratios:  \citet{AliDib14} recently showed that the faster diffusion of water compared to CO interior to the water snowline results in an increased C/O ratio in the inner disk. In general diffusion rates depend on concentration gradients, and are therefore especially efficient for species X close to the X condensation line.  The level of C/H excess in the disk gas interior to the CO snow-line will therefore depend on how far the pebbles drift inward of the CO snowline before sublimating, on the diffusion efficiency, and on the fraction of CO vapor that following back-diffusion condenses out on grains that grow and drift back into the sublimation zone. The relative rates of these different processes are likely time dependent and thus the level of achieved C/H excess in the inner disk will vary across both space and time. In the following section we use a small set of toy models to evaluate possible levels of C/H enrichments interior to the CO snowline when pebble drift redistribute CO from the outer to the inner disk and diffusion spreads out the sublimated CO vapor.

\section{Toy Models  \label{sec:res}}

\begin{figure*}[ht!]
\figurenum{2}
\plotone{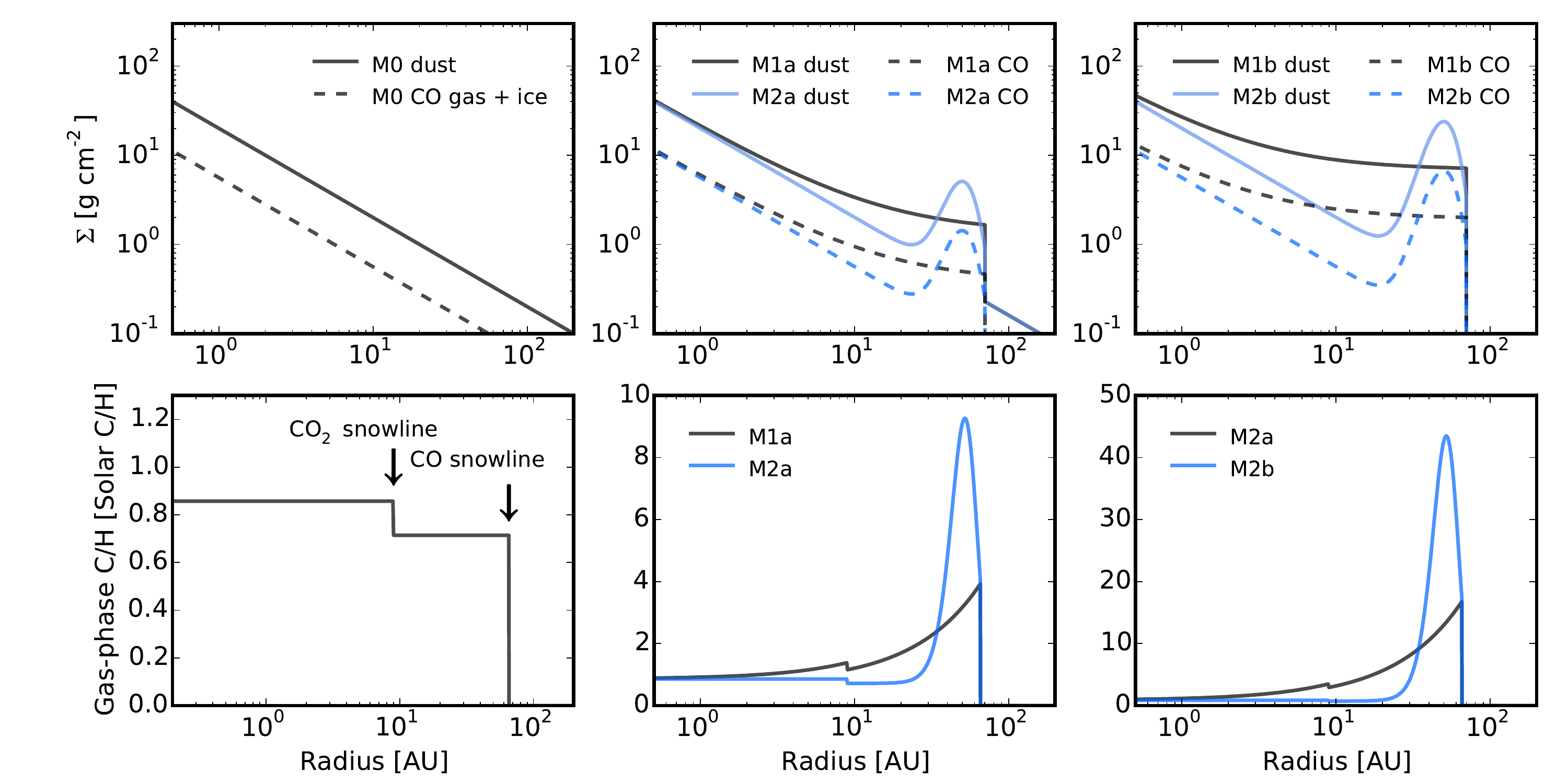}
\caption{Dust and total (solid+gas) CO distributions (upper panels), and gas-phase C/H (lower panels) in a static MMSN-type disk (left panels), and in toy models of redistribution of cold pebbles from the outer disk (exterior to 70~AU) to the inner disk due to drift. Because CO is frozen out exterior to 65~AU the CO dust re-distribution results in an increase in CO (dust+gas) in the inner disk, which interior to the CO snowline enhance gas-phase C/H. Model 1 assumes that 20\% of the outer disk solids drift inwards, while Model 2 assumes an almost complete, 99\% redistribution. CO is deposited with a Gaussian distribution of increasing width in the 'a', 'b', and 'c' models \label{fig:model}}
\end{figure*}

Figure \ref{fig:model} shows our disk toy models: the adopted initial dust (refractory solids) and CO distributions, different redistributions of solids from the outer disk, and their impacts on C/H gas-phase ratios assuming different levels of CO diffusion. Our initial disk model is a recent variation of the Minimum Mass Solar Nebula (MMSN) density and temperature profile. We assume a gas surface density of $2000\times r^{-1}$~g~cm$^{-2}$ and temperature of $120\times r^{-3/7}$~K \citep{Chiang10} between 0.1 and 200~AU. Initially the dust follow the gas distribution with a mass ratio of 1/100 (not including volatile mantles). The initial CO column density is set such that the total (solid/ice+gas) CO abundance in number of molecules per H nuclei is $1.5\times10^{-4}$ at each disk radius, corresponding to a CO/dust mass ratio of  0.42. This is shown in the upper left panel of Fig.  \ref{fig:model}, labeled M0.

To calculate snowline locations and C/H gas ratios in the disk we adopt the same silicate, carbon grain, H$_2$O, CO$_2$ and CO abundances and sublimation temperatures as in \cite{Oberg11e}, except for a slightly elevated CO sublimation temperature of 25~K. The resulting H$_2$O, CO$_2$ and CO midplane snowline locations are at $\sim$0.7, 9 and 39~AU respectively. Across the disk the C/H ratio is calculated by a balance between freeze-out and sublimation of these species at each disk radius. The lower left panel in Figure \ref{fig:model} shows the expected drops in C/H gas-phase abundances at the CO$_2$ and CO snowlines in the static disk, where substantial amounts of carbon become depleted into the solids. C/H is $\sim70$\% of the Solar value in between the CO$_2$ and CO snowlines. 

We then simulate the effects of dust grain growth and drift on the dust surface density and further on the total CO surface density by moving a fraction of the solids located exterior to 40 AU into the inner disk. The 40~AU radius is somewhat arbitrary -- the sizes of observed pebble disks range from 15 of AU and upward \citep{Tazzari16}. We opt to put it close to the CO snowline, since sublimation lines are proposed to cause disk sub-structure  \citep[e.g.][]{Zhang15}, and thus to slow down drift locally. The fraction of redistributed solids is either 20\% (Model~1/M1), appropriate at `early' times, or 99\% (Model~2/M2), characteristic of mature disks where the outer disks have been emptied of pebbles and only micron-sized grains remain. For simplicity, the pebbles are re-distributed such that a constant surface density of refractory grain mass are added to each disk radius interior to 40~AU (Fig. \ref{fig:model} top panels) -- the adopted redistributed dust profile have a little effect on the output C/H ratio. The gas-solid balance of water and CO$_2$ is then recalculated using the new surface density of solids assuming a constant H$_2$O/CO$_2$/refractory solids ratio throughout the disk. This is a reasonable assumption exterior to the CO$_2$ snowline (the region of interest for this paper), since the pebbles originate beyond the CO snowline and thus all available H$_2$O, CO$_2$ and CO in the outer disk are brought with the pebbles as they drift. 

We treat CO differently than water and CO$_2$, since the drifting pebbles and CO decouple where CO mantles sublimate, which happens somewhere between the static disk CO snowline of 39~AU and $\sim$20~AU dependent on the pebble size and drift speed \citep{Piso15}. We adopt a 30~AU sublimation radius and then spread the CO vapor inward and outward, simulating the effect of diffusion \citep[e.g.][]{Ros13,AliDib14,Owen14}. The diffusion efficiency is not well known and probably varies between disks, and during disk lifetimes. In the toy models we spread the sublimated CO using Gaussian distributions of the CO surface density with widths $\sigma$=5, 10 and 20~AU, labeled 'a', 'b', and 'c' in Fig. \ref{fig:model}, assuming for simplicity that forward and backward diffusion are symmetric. These toy model redistributions allow us to explore the limiting cases of almost no vs. efficient CO gas diffusion following sublimation. 

Following re-calculations of the total H$_2$O, CO$_2$ and CO column densities in the disk, we use the same condensation-sublimation steady-state calculations as outlined above to determine their divisions between gas and solid across the disk. In reality the sublimation-condensation steady-state will not be reached instantaneously as assumed here, and the time scale depends on the density of grain surface area. In our second redistribution scenario, where the outer disk is almost emptied of solids, the condensation timescale of CO will be substantially longer than standard assumptions. The increase is not as high as might be imagined, however, since grains that remain in outer disk regions are  small, based on observations of lack of millimeter emission at large disk radii in disks that show sign of pebble drift, and such grains provide much surface per solid mass density. The remaining panels in Fig. \ref{fig:model} show the resulting C/H ratios in our toy models of dust and CO redistributions following pebble drift. The C/H ratio reaches supersolar values between the CO$_2$ and CO snowlines in all toy models except for the scenario with efficient diffusion and inefficient pebble drift. The most `optimistic' scenario, i.e. a disk where almost all dust exterior to 40~AU form pebbles and drift inward and deposit all CO around 30~AU results in a maximum C/H ratio of more than 10x Solar between the CO$_2$ and CO snowlines.

\begin{figure}[ht!]
\figurenum{3}
\plotone{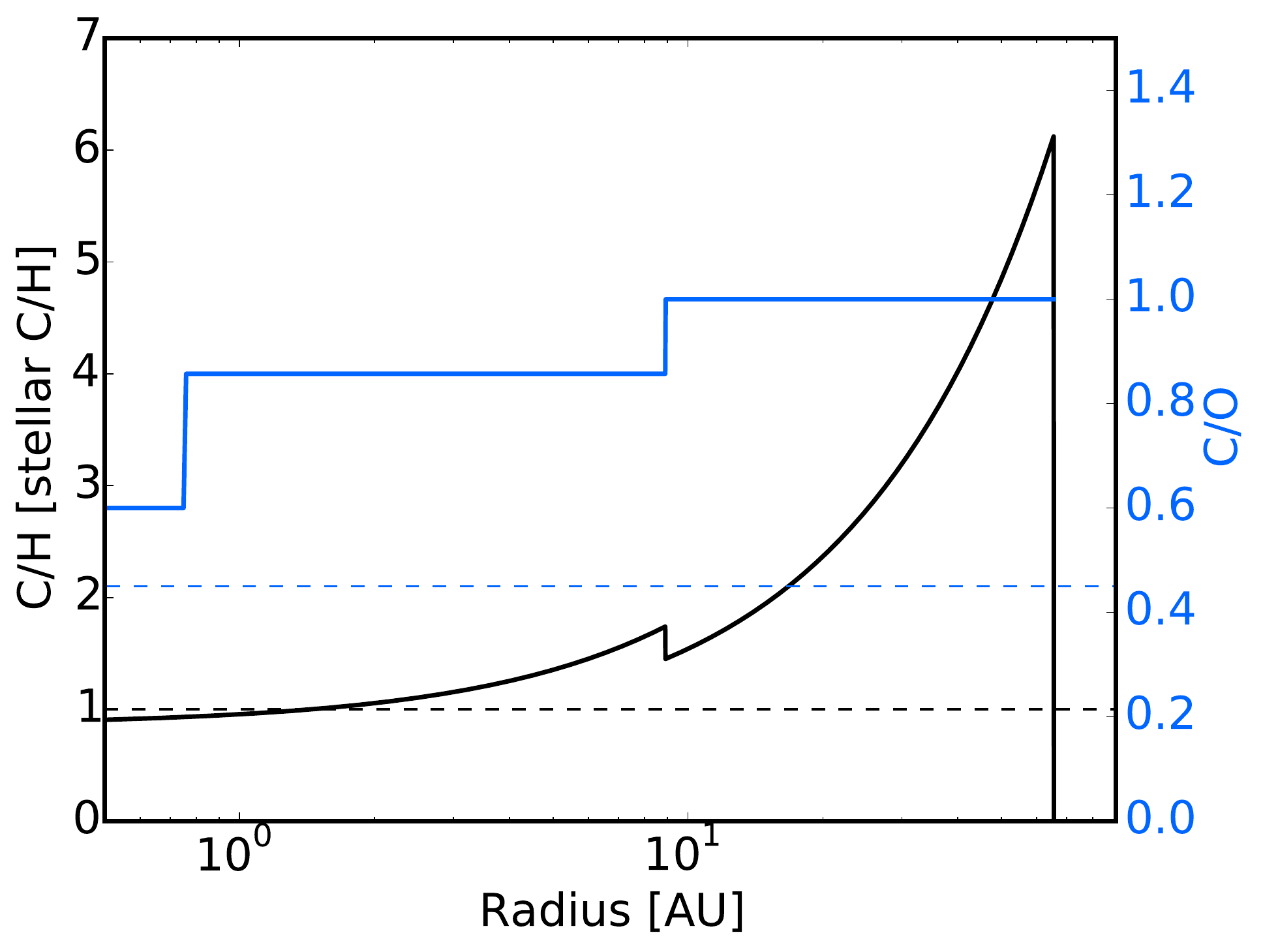}
\caption{C/H and C/O gas-phase ratios in a disk where the inner disk has been enriched in CO following drift and sublimation of grains initially located exterior to the CO snowline. The jumps in C/O at 0.7 and 9~AU are caused by water and CO$_2$ snowlines, while the drop in gas-phase C/H at 65~AU is caused by the CO snowline. \label{fig:coh}}
\end{figure}

Figure \ref{fig:coh} shows C/H and C/O in a disk assuming intermediate drift and CO diffusion efficiencies, i.e. where 50\% of the outer disk solids have been redistributed and CO vapor diffusion spreads the sublimated CO 10~AU in each direction. The C/O pattern is similar to \citet{Oberg11e}. In the region where gas-phase C/O is unity, the C/H is enhanced by up to a factor of three compared to the Solar value. Snowlines + pebble drift can thus cause co-located C/O and C/H enhancements in disks beyond the water and CO$_2$ snowlines. This is in contrast compared to previous predictions based on a static disk model.

\section{Discussion}

In a previous study we suggested that planetary atmospheres can contain excess carbon with respect to oxygen, i.e. an enhanced C/O ratio, when they mainly form from oxygen depleted gas \citep{Oberg11e}.  An oxygen depleted gas in the outer regions of protoplanetary disks is a natural outcome of sequential freeze-out of oxygen and carbon-rich volatiles. Thus carbon rich Gas Giants forming in the outer disk should appear enriched in C/O, but depleted in C/H and O/H. This pattern of enhanced C/O and depleted C/H was proposed as distinguishing feature between Gas Giant formation in the inner and outer regions of protoplanetary disks; in the inner disk planet atmospheres could become enriched  in C over O if polluted by carbon-rich planetesimals, which should result in superstellar C/H \citep{Lodders04}.  The results from our toy model demonstrates that when considering that disks are dynamic objects, this simple distinction breaks down. Since pebble drift should be universal, we rather expect that Gas Giant envelopes accreted exterior to the water snowline are rich both in C/O and C/H. 

This result raises the question whether there are other compositional features in the atmospheres of extra-Solar Gas Giants that could be used to distinguish between different Gas Giant envelope formation locations. We propose that the N/C ratio in conjunction with C/O and C/H may provide such a test. Inner Solar System solids, as traced by Earth mantle composition and meteorites, have low average N/C ratios, perhaps as low as 0.02 based on the Earth's mantle \citep{Bergin15}; the Solar value is an order of magnitude higher, i.e. N/C=0.28 \citep{Lodders03}. By contrast the gas-phase N/C in the outer disk should be supersolar in a static or steady state disk, because most nitrogen is bound up in the volatile N$_2$, resulting in little nitrogen depletion in the gas compared to carbon and oxygen (Piso et al. ApJ subm.). The relative amounts of CO and N$_2$ that will be redistributed in the inner disk due to drift, and thus the exact N/C ratio, will depend on the details of the disk structure, especially the exact locations of CO and N$_2$ snowlines. 

Gas Giant envelope compositions will not only depend on factors considered in the toy models, (pebble drift, CO and N$_2$ sublimation and diffusion), however but also on pollution by solids through core dredging and pebble and planetesimal accretion and sublimation in Gas Giant atmospheres. All these `polluting'  processes are bound to happen to some degree, but the effects are difficult to predict, both qualitatively and quantitatively, because of interplay between migration, pebble drift and Gas Giant internal structure and dynamics. In general solid `pollution' of the envelope is expected to reduce the C/O ratio, since solids are typically oxygen-rich, and increase the C/H ratio, since even oxygen-rich solids always contain some carbon. Nascent planets can sample solids from many different disk regions throughout its formation because of migration \citep{Goldreich80,Lin93}. This implies that the exact C/O ratio of polluting solids is difficult to predict and will be different if the main pollution channel is core dredging, pebble or later solid accretion. By contrast, the nascent gas envelope should sample a more limited range of disk environments and C/O and C/H ratios, since runaway accretion of gas is fast compared to migration.

Finally, our model result provides an alternative interpretive framework for C and N enrichments in Jupiter and Saturn. It is generally assumed that their supersolar C/H and N/H ratios are the result of the accretion of  icy planetesimals \citep{Owen99}. While this must happen at some level, the amount of potential carbon and nitrogen mass within the population of drifting pebbles from the outer disk may far exceed the mass within a proto-Jovian feeding zone. If these pebbles deposit their volatiles at the birth radii of Jupiter and Saturn, then we would expect to see significant enrichment of volatile elements such as C and N in their gaseous envelopes from accretion of gas alone. This may explain why C and N are similarly enhanced in Jupiter, which is not expected from planetesimal accretion \citep{Wong04}. In addition to explaining carbon and nitrogen enhancements, this model may cast a new light on the noble gas composition of Jupiter and Saturn, whose origin is currently debated \citep[e.g.][]{Wilson10,Nikhil15}. The same pebbles that brought CO and N$_2$-rich into the inner Solar System would also have brought with them noble gases -- several of the noble gases have sublimation temperatures similar to CO and N$_2$ \citep[][Fayolle private communication]{Schlichting92,Fayolle16}. This could explain why Ar, Kr and Xe are similarly enhanced to C and N in the atmosphere of Jupiter \citep{Niemann98}. 

In conclusion adding grain drift and solids and volatile re-distribution to disk models will increase the C/H ratio in the disk regions where C/O is unity due to sequential freeze-out of carbon and oxygen bearing volatiles. This implies that we should expect to find large separation Gas Giants with both superstellar C/O and C/H. To conclusively distinguish such planets from Gas Giants forming in the inner disk, where C/O enhancements may occur because of accretion of carbon-rich planetesimals, would require additional measurements of e.g. nitrogen or noble gases, or constraints on the dynamical history of the planet.


\acknowledgments

The authors are grateful to Ellen Price for computational advise, and for helpful comments from an anonymous referee. KI\"O acknowledges
funding through a Packard Fellowship for Science and
Engineering from the David and Lucile Packard Foundation. EAB acknowledges support from the National Science Foundation under grants AST-1514670 and AST-1344133 (INSPIRE),
and NASA XRP program NNX16AB48G


\bibliographystyle{aasjournal}

\end{document}